\documentclass[aps,reprint]{revtex4-1}

\usepackage{graphicx}
\usepackage{amsmath,amssymb}

\draft 

\begin{document}

\title{Reduction of spin polarization by incoherent tunneling in Co$_2$FeAl/MgO/CoFe magnetic tunnel junctions with thick MgO barriers}

\author{M.S. Gabor}
\email[Electronic mail: ]{Mihai.Gabor@phys.utcluj.ro}
\affiliation{Centre for Superconductivity, Spintronics and Surface Science (C4S), Technical University of Cluj-Napoca, Cluj-Napoca, Romania}
\affiliation{Institut Jean Lamour, UMR 7198, CNRS - Lorraine University, Nancy, France}

\author{C. Tiusan}
\email[Electronic mail: ]{Coriolan.Tiusan@phys.utcluj.ro}
\affiliation{Centre for Superconductivity, Spintronics and Surface Science (C4S), Technical University of Cluj-Napoca, Cluj-Napoca, Romania}
\affiliation{Institut Jean Lamour, UMR 7198, CNRS - Lorraine University, Nancy, France}

\author{T. Petrisor Jr.}
\affiliation{Centre for Superconductivity, Spintronics and Surface Science (C4S), Technical University of Cluj-Napoca, Cluj-Napoca, Romania}
\affiliation{Institut Jean Lamour, UMR 7198, CNRS - Lorraine University, Nancy, France}

\author{T. Petrisor}
\affiliation{Centre for Superconductivity, Spintronics and Surface Science (C4S), Technical University of Cluj-Napoca, Cluj-Napoca, Romania}

\author{M. Hehn}
\affiliation{Institut Jean Lamour, UMR 7198, CNRS - Lorraine University, Nancy, France}

\author{Y. Lu}
\affiliation{Institut Jean Lamour, UMR 7198, CNRS - Lorraine University, Nancy, France}

\author{E. Snoeck}
\affiliation{CEMES, CNRS 29 rue J. Marvig 31055 Toulouse, France}

\date{\today}

\begin{abstract}

We report on spin polarization reduction  by incoherent tunneling in realistic single crystal
Co$_2$FeAl/MgO/Co$_{50}$Fe$_{50}$ magnetic tunnel junctions (MTJ) compared to reference Fe/MgO/Fe.
A large density of misfit dislocations in the Heusler based MTJs has been insured by a thick MgO barrier and its 3.8\% lattice mismatch with the Co$_2$FeAl electrode.
Our analysis implicates a correlated structural-transport approach. The crystallographic coherence, in the real space, is investigated using High Resolution Transmission Electron Microscopy phase analysis. The electronic transport experiments in variable temperature, fitted with a theoretical extended-Glazman-Matveev model, address different levels of the tunneling mechanisms from direct to multi-center hopping.  We demonstrate a double negative impact of dislocations, as extended defects, on the tunneling polarization. Firstly, the breaking of the crystal symmetry destroys the longitudinal and lateral coherence of the propagating Bloch functions. This affects the symmetry filtering efficiency of the $\Delta_1$ states across the (100) MgO barriers and reduces the associated effective tunneling polarization. Secondly, dislocations provide localized states within the MgO gap. This determines temperature activated spin-conserving inelastic tunneling through chains of defects which are responsible for the one order of magnitude drop of the tunnel magnetoresistance from low to room temperature.

\end{abstract}


\maketitle

\section{Introduction}
\label{Introduction}

The performance of spintronic devices based on magnetic tunnel junctions (MTJs) is mainly dependent on their tunnel magnetoresistive (TMR) response amplitude. In epitaxial magnetic tunnel junctions, the TMR is a consequence of several factors: the spin polarization of the ferromagnetic electrodes, the symmetry filtering properties of the electrodes and the symmetry dependent attenuation rate of the evanescent wave function in the tunnel barrier. Consequently, two alternatives are employed in order to increase the TMR response of a MTJ. One is to use half-metallic ferromagnets (HMFs) as magnetic electrodes in MTJ. Since they have an energy gap around the Fermi level E$_F$ in the minority spin band, theoretically, they are expected to provide 100\% spin polarization. The other alternative is to use symmetry dependent half metallicty of $\Delta_1$ band in (100)bcc metals \cite{Yuasa-NatMat2004} and alloys, \cite{Parkin2004,Djayaprawira2005} which in conjunction with an epitaxial (100)MgO barrier, providing symmetry dependent atenuation rates, behave like half metals in the coherent tunneling regime \cite{Mathon2001,Butler2001}. Among the HMFs, a special class is represented by the full-Heusler alloys. Theoretical predictions \cite{Galanakis2002,Picozzi2002} indicate that the Co-based full-Heusler alloys should behave like half-metals even at room temperature. Currently, one of the most studied full-Heusler alloy is Co$_2$FeAl (CFA). It was demonstrated to provide giant tunneling magnetoresistance (GTMR) in MgO based MTJs  due to the mixed effect of large spin polarization of electrodes and $\Delta_1$ band symmetry filtering \cite{Wang2009,Wang2010,Wang2010B,texturedCFA} provided by the symmetry dependent attenuation rate of the barrier. Moreover, this material is of special importance since it was shown to have a low Gilbert damping \cite{Mizukami2009,GaborJAP}, essential for applications concerning magnetization reversal by spin torque using low current densities and for building high efficiency spin torque oscillators.

In this paper we address a special tunneling transport regime in CFA Heusler based MTJs, corresponding to thick MgO barriers, regime where the density of structural defects within the insulator is particularly important. We show the negative effect of extended defects within the barrier, e.g. dislocations and strain fields, on the effective tunneling spin polarization. Our experimental strategy combines detailed structural analysis correlated with tunneling transport in variable field and temperature. We demonstrate that, due to the large density of dislocations and their specific extended geometry relative to the barrier thickness, the lateral and the longitudinal coherence of the propagating wave function is lost within the barrier. The structural defects in the MTJ stack, such as dislocations and strain fields, are clearly identified and analyzed in terms of geometry and density using the 'Geometrical Phase Analysis' \cite{Snoeck} of cross section High Resolution Transmission Electron Microscopy (HRTEM) images. This structural analysis is correlated with specific magneto-transport experiments in variable temperature treated within the theoretical extended Glazman-Matveev model\cite{Lu}. This transport analysis allowed us to demonstrate that spin-conserving inelastic tunneling through chains of localized states associated with extended spatial defects within the tunneling barriers is dominant, detrimental to the direct tunneling mechanism expected to provide large effective tunneling polarization.

\section{Results and Discussions}

MTJs with the structure MgO(001)//Cr(20nm)/ CFA(40nm)/ MgO(4.2nm)/ CoFe(10nm)/ Co(40nm)/ Au(10nm), have been grown by combining two deposition techniques: sputtering and Molecular Beam Epitaxy (MBE). The Cr buffer layer and bottom CFA electrode have been grown in a magnetron sputtering system, as described elsewhere \cite{GaborPRB}. Afterwards, the MgO//Cr/CFA stack has been transferred in an MBE chamber equipped with Reflection High Energy Electron Diffraction (RHEED) \textit{in-situ} analysis. Fig. 1a shows the RHEED pattern of the as deposited CFA film. The shape of the pattern demonstrated that film is crystalline but rough within the RHEED coherence scale-length. Moreover, one can remark a diffusive background most probably related to the chemical disorder present in the as-deposited sample.
\begin{figure}
\begin{center}
\includegraphics[width=1\linewidth]{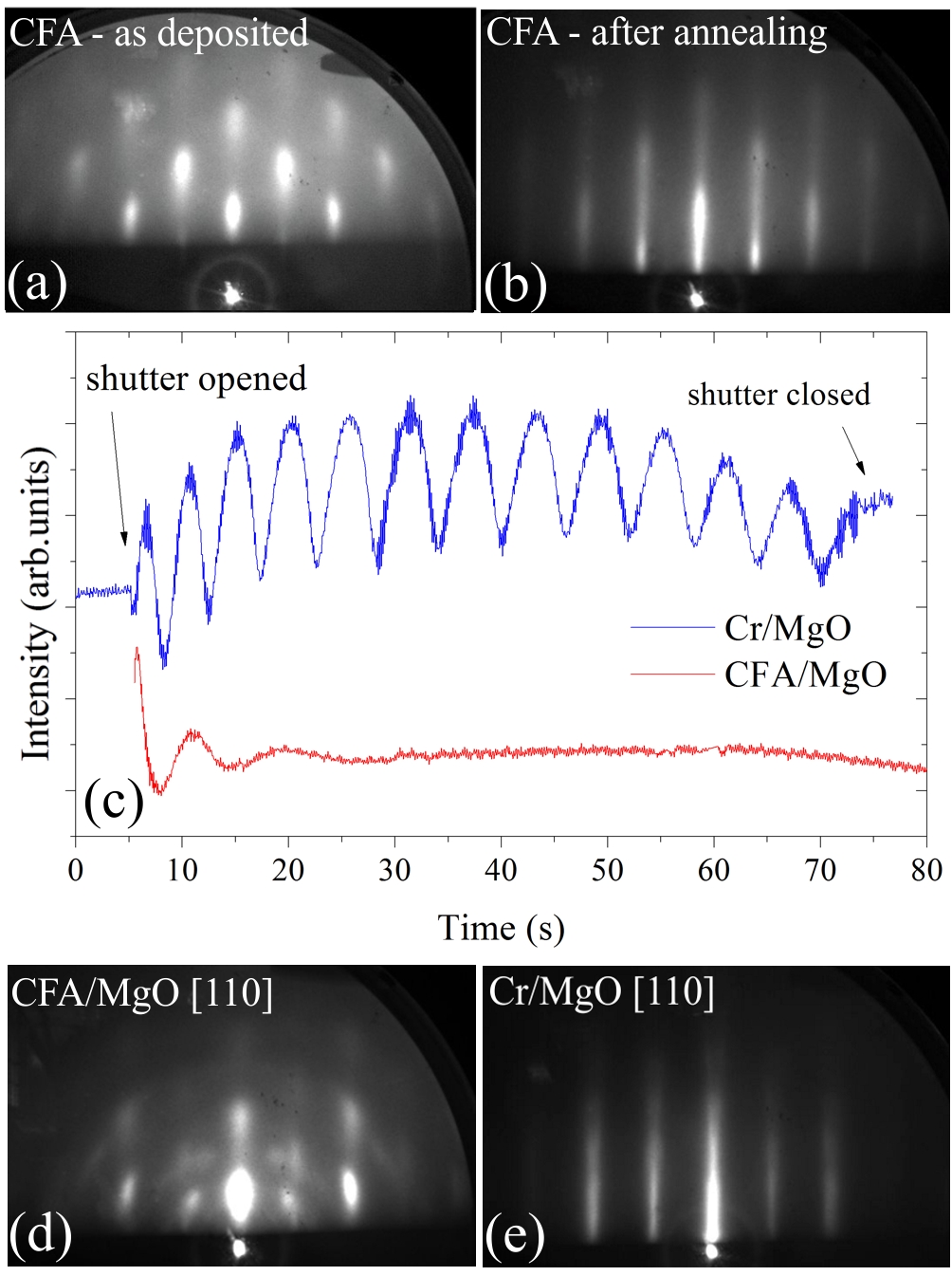}
\caption{\label{fig1} RHEED patterns of (a) as deposited CFA film and (b) annealed at 600$^\circ$C; (c) RHEED intensity oscillations for a MgO film grown on CFA and Cr; RHEED patterns of the MgO film after growth on (d) CFA and (e) Cr.}
\end{center}
\end{figure}
After annealing at 600$^\circ$C for 20 minutes, the CFA layer is flattened (Fig. 1b), the reduction of the diffusive background being correlated with the improvement of the chemical ordering of the system from A2 towards the B2 phase \cite{GaborPRB}. On top of this CFA layer, after cooling down to 120$^\circ$C, the MgO barrier has been grown by electron beam evaporation within the MBE chamber. During the barreir growth, the pressure in the deposition chamber increased from $5\times10^{-11}$ Torr to $10^{-8}$ Torr due to the enhancement of oxygen partial pressure as confirmed by in-situ Quadrupole Mass Spectrometry analysis.

The growth kinetics of MgO on CFA has been carefully monitored by RHEED. Analysis of intensity oscillations (Fig. 1c) demonstrates a 3D growth mode of MgO on CFA. The oscillations are damped rapidly, in contrast with the growth of MgO on (001) Cr (Fig. 1c) where a clear layer-by-layer growth is asserted by the RHEED intensity oscillations with a low attenuation. Moreover, the 3D growth of MgO on CFA is confirmed by the final RHEED pattern of a 4.2 nm thick MgO barrier (Fig. 1d). We observe that the MgO exhibits a poorly (001) crystallized structure mixed with a polycrystalline phase and a rough surface within the RHEED coherence scale, probably related to the complex chemical structure of CFA surface in B2 phase (random distribution of Fe and Al atoms within their atomic sites). In contrast, similar RHEED pattern recorded for a 12ML MgO grown on Cr (reference sample for MgO layer-by-layer growth), corresponding to the RHEED oscillations in Fig. 1c, show that in this case the barrier grown on Cr is flat and with an almost perfect (001) orientation. The RHEED analysis of the barrier during its growth provides a first indication about the poor structural quality of the insulator. This will be furthermore confirmed by the more detailed HRTEM analysis performed on the complete MTJ stack.

After the barrier deposition, the top CoFe magnetic electrode of the magnetic tunnel junction has been grown by MBE at room temperature (RT). As expected from the poor crystal quality of the MgO barrier, the as-deposited CoFe layer consists of a (001) crystallized structure mixed with a polycrystalline phase (Fig. 2a). However, an in-situ annealing of the CFA/MgO/CoFe stack at 450$^\circ$C promotes layers crystallization improving also the surface flatness (Fig.2b), as confirmed also by the x-ray 2~theta-theta diffraction pattern (Fig. 2c). To operate magnetically the tunnel junction, a hard-soft architecture is required. In our MTJ stack, the magnetic hardening of the top CoFe layer is obtained by direct exchange coupling with an additional 40nm thick Co film epitaxially grown on CoFe in a hexagonal phase with the c axis parallel to the film plane \cite{Faure-Vincent}.

One can expect that the annealing of the top CoFe electrode will also affect the barrier quality, previously analyzed by RHEED. To confirm this, an evaluation on the structural quality of the whole stack was performed by HRTEM analysis on a cross-sectional specimen.

\begin{figure}
\begin{center}
\includegraphics[width=1\linewidth]{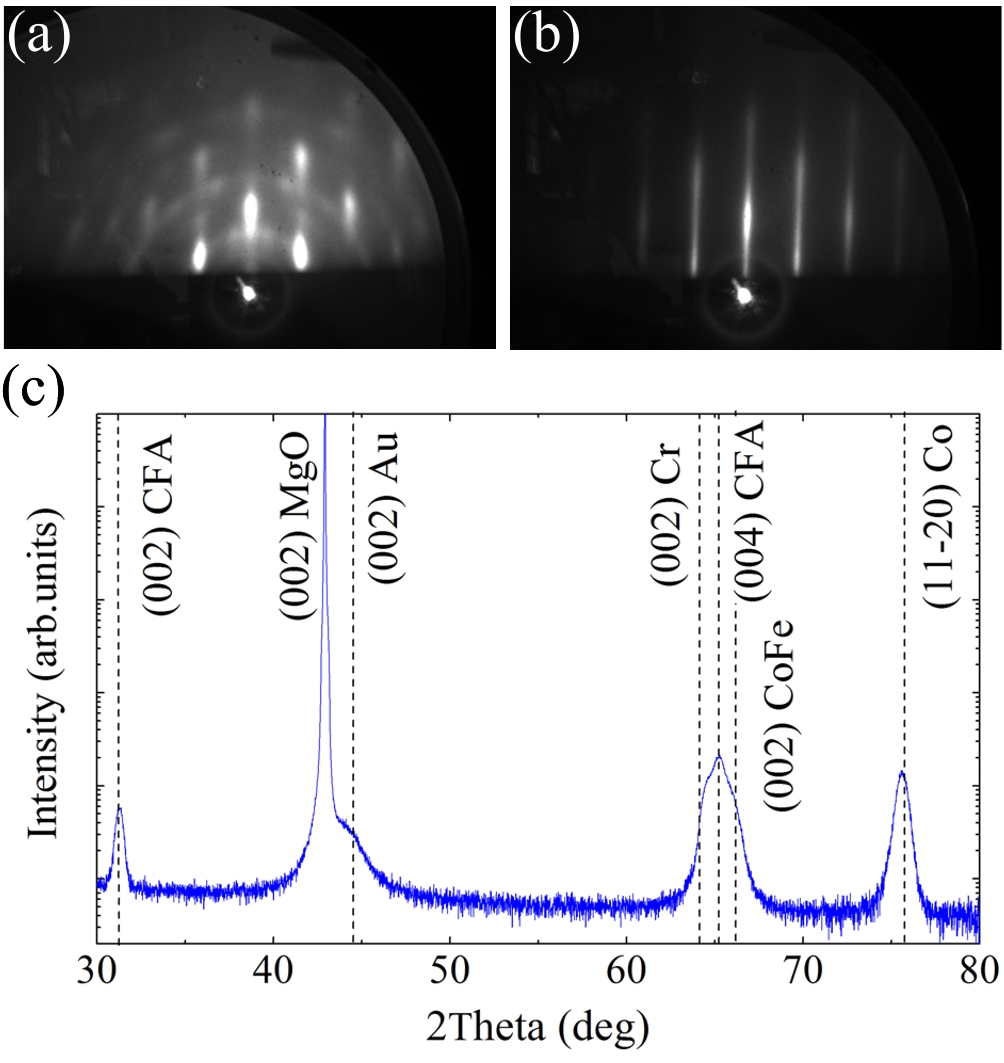}
\caption{\label{fig2} RHEED patterns for the CoFe electrode (a) as deposited and (b) annealed at 450$^\circ$C; (c) x-ray 2Theta-Theta diffraction pattern of the complete MTJ stack.}
\end{center}
\end{figure}

At a first glance, the HRTEM images (Fig. 3a) demonstrate a good crystalline quality and relative flat interfaces. This confirms that the annealing stage of the upper CoFe electrode triggers the crystallization of the MgO tunnel barrier and improves the quality of the interfaces. However, accurate analysis of the interface roughness from HRTEM is delicate because it integrates the roughness profile over the analyzed section width. Measurements of the atomic plane spacing show that the MgO barrier is totally relaxed, d(002) being equal to d(200) and with a value of 0.21 nm, which is expected for a bulk MgO crystal. The lattice mismatch between CFA and MgO is about 3.8\%, which gives considerable strain that relaxes through the formation of misfit dislocations. The plastic relaxation of MgO on CFA takes place in the early stages of growth, which explains the rapid oscillation damping of the \textit{in-situ} RHEED intensity oscillations (Fig. 1c).

Detailed information about structural defects such as dislocations and local strain fields can be extracted from HRTEM by performing the so-called "Geometrical Phase Analysis" (GPA) procedure \cite{Snoeck}  (Fig. 3b,c,d). The GPA method consists on performing a Fourier Transform (FT) of a HRTEM image and selecting in its power spectrum reflections corresponding to lattice periodicities which present spatial deviations. Any spatial deviations in the real space correspond to phase shift in the reciprocal space and the latter can be much more easily highlighted. As an example, when performing the inverse Fourier Transform (IFT) of the HRTEM image in Fig 3a and selecting the 002 reflection perpendicular to the interfaces, the lattice fringes corresponding to the (200) planes can be highlighted and discontinuities in these planes appear (arrows in Fig 3b). They correspond to misfit dislocations. These misfit dislocations can be even better seen when calculating the phase shift of the (200) plane periodicities between regions of different structures i.e. of different spatial periodicities (Fig. 3c). The dislocations are perpendicular to the CFA/MgO interfaces with a periodicity of approximately 4.4 nm and correlated from one side of the barrier to the other. This becomes even clearer from the Fig. 7b which shows the missing planes within the insulating barrier. One can see that the dislocations from the CFA electrode continue throughout the barrier and to the upper electrode. Inside the barrier, among the misfit dislocations that are relatively perpendicular to the interfaces, one can observe dislocations that impinge on the interface at much smaller angles and continue over tens of nanometers. When performing the local derivative of the phase image, the relative deformation of a given lattice compared to the other one can be measured (Fig. 3d). [16] Figure 3d shows the local deformation  $\varepsilon_{yy} = du_{yy}/dy$ in the direction parallel to the interface. The mean deformation value within the interface (4\%) corresponds to the misfit between MgO and CFA. Figs. 3c and 3d indicate that the deformation correlated with the misfit dislocations is located in regions of about 1.5 to 2 nm width at each interface. From the HRTEM analysis an important conclusion can be drawn. Even though the crystalline quality of the barrier increases, after the annealing of the upper CoFe electrode, it contains a large density of defects, like lattice deformations and dislocations. That, as we will demonstrate in this paper, will affect the spin dependent tunneling process.

\begin{figure}
\begin{center}
\includegraphics[width=1\linewidth]{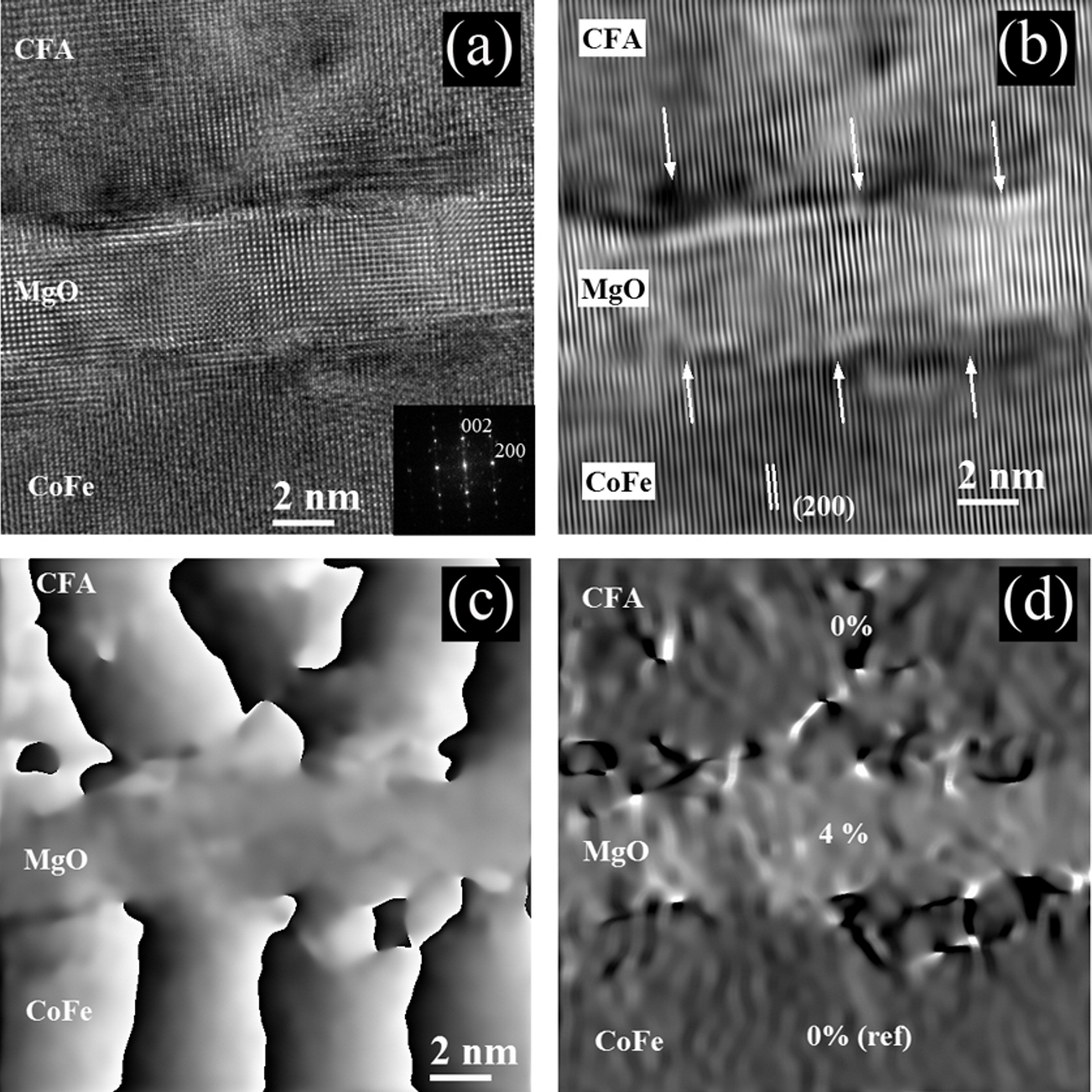}
\caption{\label{fig3} (a) HRTEM image of the CFA/MgO/CoFe trilayer with its FT inset. (b) (200) filtered image showing the discontinuities of the (200) planes across the interfaces (arrows). (c) Phase image evidencing the additional planes in the two electrodes and the associated misfit dislocations and (d) local deformation  $\varepsilon_{yy} = du_{yy}/dy$ image. }
\end{center}
\end{figure}

To perform magneto-transport experiments, our MTJ stacks have been patterned by UV lithography and ion beam etching in square MTJ devices with lateral size from 10 to 50 $\mu m$. In Fig. 4 we illustrate a typical  tunnel magnetoresistance (TMR) curves for the CFA/ MgO/CoFe MTJ measured both at low (14K) and at room temperature (300K). The first striking remark is the low TMR ratio at room temperature (2.8\%) measured in our systems. This is in significant contrast to standard results obtained in similar systems, but with lower barrier thickness, reported in literature (larger than 300\% ratios at RT in CFA/MgO single crystal MTJs, 166\% at RT in textured CFA/MgO MTJs \cite{Wang2009,Wang2010,Wang2010B,texturedCFA}). The room-temperature (RT) TMR increases one order of magnitude, to about 23\%, when the temperature is lowered.
Following the HRTEM analysis, a first assumption which could be driven is that the structural defects (strain field, dislocation) should affect the spin polarized tunneling mechanisms and would have negative effects of tunneling polarization amplitude.

\begin{figure}
\begin{center}
\includegraphics[width=1\linewidth]{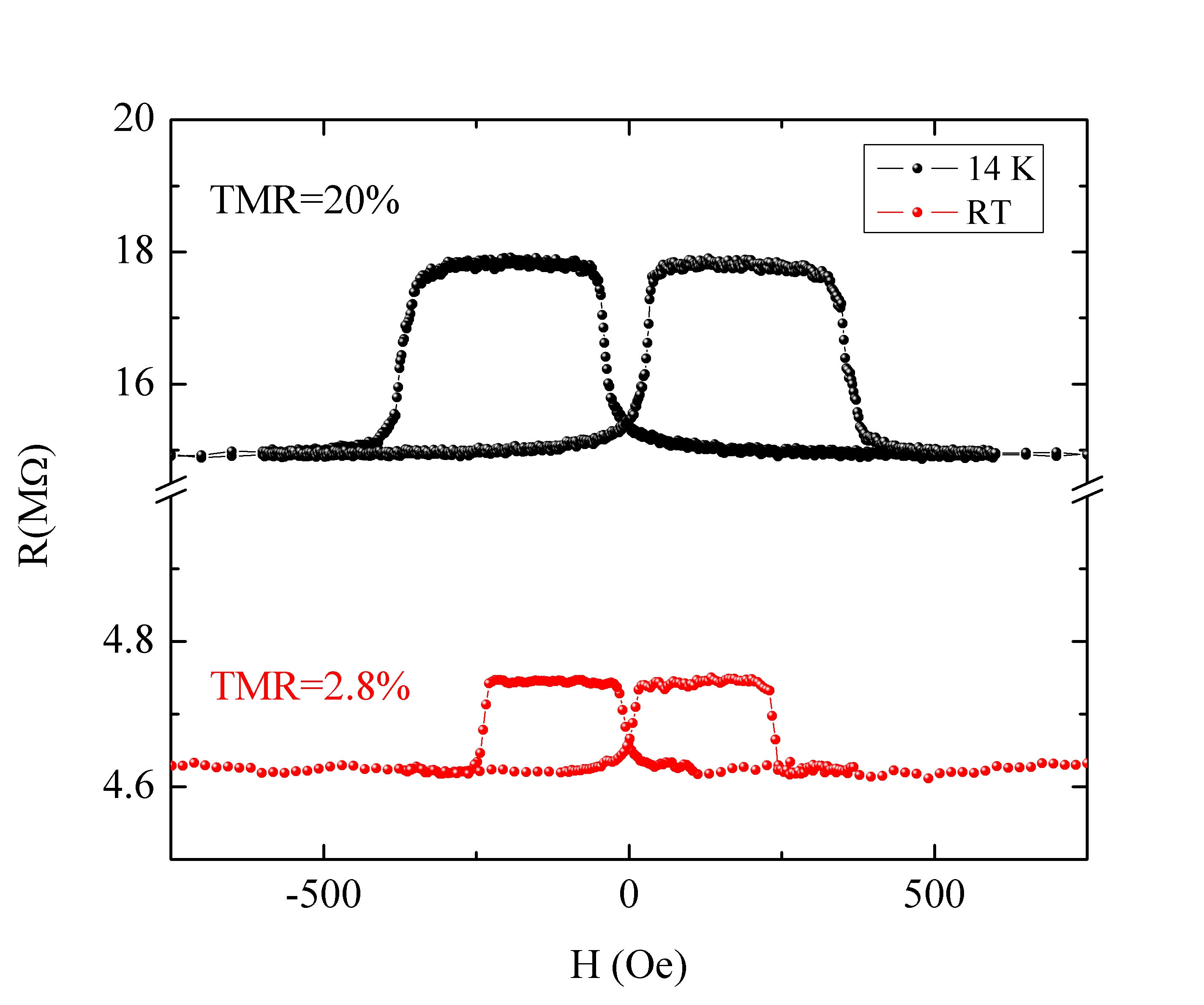}
\caption{\label{fig4} (a) Magnetoresistance measurements performed on the MTJs at low (14 K) and room temperature (300 K)}
\end{center}
\end{figure}

To support this assumption, drawn from  HRTEM and to provide deeper understanding on the origin of the low tunneling polarization ratio, unusual for single crystal MTJ devices, we performed tunnel transport experiments in variable temperature. This kind of analysis is commonly used to address the transport mechanisms in magnetic tunnel junctions. Our specific analysis has been oriented to investigate the complex transport channels contributing to the tunneling transport and it is performed within the framework of the extended Glazman-Matveev (GM) model \cite{Lu} of spin-conserving hopping processes through chains of N localized states (LS) induced by the defects already pointed out by HRTEM.

Within this GM model the conduction in the antiparallel (AP) magnetization configuration has form:
$
G_{AP} = \sigma_0 + \sum \sigma_N T^{\nu_N},
$
where $\sigma_N$ describes the hopping through $N\geq2$ LS, $\sigma_0$ stands for the direct and 1-LS hopping and $\nu_N= N - 2/(N+1)$ is a characteristic exponent. Figure 5a shows the dependence on the temperature of the AP conductance as well as a very good fit of the experimental points using the above equation. A first strong remark is that, in order to fit correctly the data, up to 6-LS chains were needed. The fit was done using only up to 2-LS chains in the low temperature range regime. Then, progressively with increasing the temperature range, higher order LS chains have been introduced when required, until the whole temperature range was perfectly fitted. Due to their different $T^{\nu_{N}}$ signature, the conduction of the different N-LS chains can be extracted from the fit. In figure 5b we have plotted the weighted contribution $W_N=G^{N}_{AP}/G_{AP}$ of each conduction channel to the total conductance in the AP state. This analysis provides important information about the tunneling contribution of different channels as a function of temperature. While, in the low temperature regime the conductance is mainly due to the 0 and 1-LS contribution (which cannot be distingueshed within the extended GM approach \cite{Lu}), when the temperature is increased towards the room temperature (RT) higher order conductivity chains ($N\geq2$) become dominant. Within the model of spin-conserving hopping \cite{Lu} the TMR assigned to a variety containing N-LS is given by $TMR(N)=\frac{(1+P)^{2\beta_N} + (1-P)^{2\beta_N}}{2(1-P^2)^{\beta_N}}-1$, where $\beta_N =1/(N+1)$ and $P$ is the effective temperature dependent spin polarization of electrodes $P=P_0(1-\alpha T^{3/2})$ \cite{Shang98}, $\alpha$ is the spin-wave parameter related to the interfacial Curie temperature and $P_0$ is the effective spin polarization at 0 K. A direct result of the above relation is that even if the channels are spin conserving, the TMR gradually decreases when N is increased. The total TMR is given by the sum of the TMR of each conduction variety weighted by their fractional contribution $TMR=\sum_N W_N(T)\times TMR(N)$.

\begin{figure}
\begin{center}
\includegraphics[width=1\linewidth]{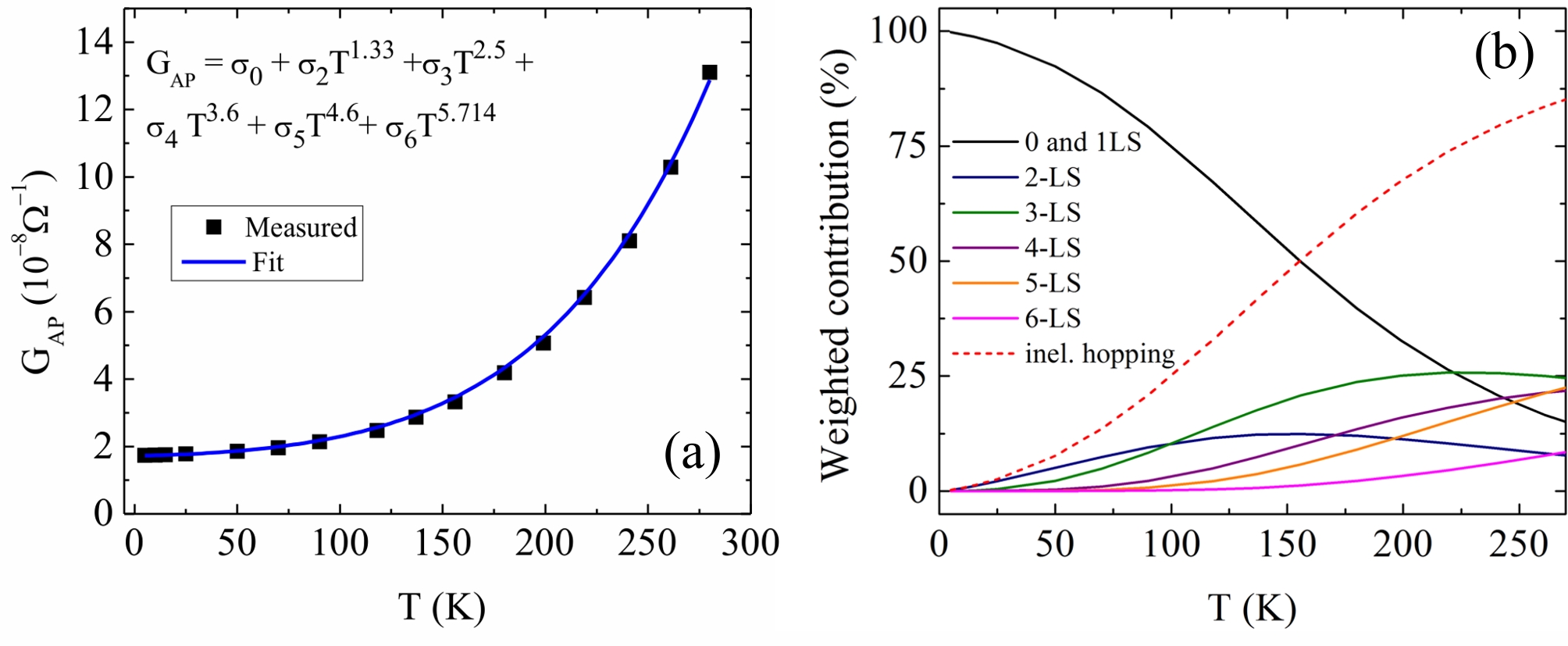}
\caption{\label{fig5} (a) Temperature dependence of the tunnel conductance in the AP state. (b) Relative contribution of the N-LS chains to the total conductance in the AP state, and relative contribution of total inelastic hopping ($N\geq2$) to the total conductance.}
\end{center}
\end{figure}

\begin{figure}
\begin{center}
\includegraphics[width=1\linewidth]{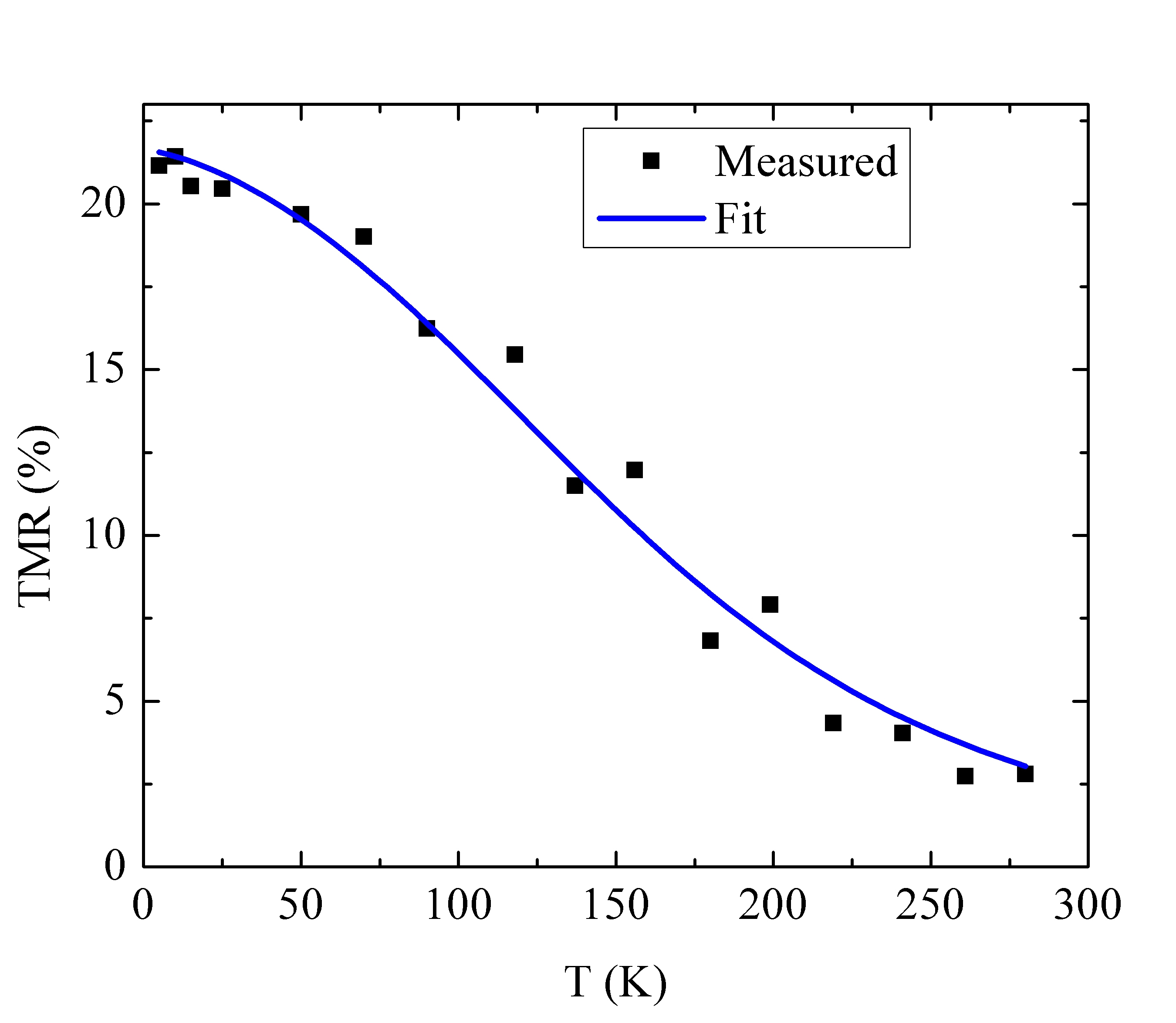}
\caption{\label{fig6} TMR dependence on temperature}
\end{center}
\end{figure}

Figure 6 shows the evolution of the TMR with temperature as well as a fit using the above described model. As can be seen from the figure, the fit reproduces well the experimentally observed strong TMR decrease with temperature. In our fitting procedure, due to the large thickness of the barrier and confronting with phase images extracted from HRTEM (see later Fig. 7b,e and related discussion), we reasonably ignored the direct channel and took into account only the resonant tunneling with one  ($N=1$) and higher order hopping channels ($N\geq2$). Consequently, our extended GM analysis provided an effective polarization of $P_0=0.57$. This is in agreement with the theoretical and experimental expected values for CFA films showing B2 disorder \cite{Miura04,Inomata08,schebaum}. Therefore, one can conclude that despite standard electrode polarization provided by our CFA films with B2 disorder, the low temperature TMR is reduced from 0.96, as expected for direct tunneling (N=0), to 0.23 due to the dominant resonant tunneling (N=1) via localized levels associated with extended structural defects within the thick insulator. Furthermore, the fact that the value of the effective spin polarization of $P_0=0.57$ is in agreement with the theoretical and experimental expected values for CFA films showing B2 disorder indicates that that our thick barrier MTJs do not benefit from the $\Delta_1$ band symmetry filtering of the MgO barrier, which should increase the effective tunneling spin polarization to higher values \cite{Wang2009,Wang2010,Wang2010B,texturedCFA}. This is a direct consequence of the dominating incoherent hopping assisted transport in our thick barrier CFA/MgO/CoFe MTJs. To further confirm this we performed detailed structural investigations by HRTEM  phase image contrast focused on the MgO barrier zone, correlated with tunneling spectroscopy experiments (dI/dV curves) measured in patterned MTJ devices. The results (Fig 7a,b,g) were compared to those obtained in standard Fe/MgO/Fe MTJs (Fig. 7c,d,h) where we have previously demonstrated that tunneling coherent spin and symmetry dependent tunneling channels dominate the conductance and are responsible on the large tunneling polarization and TMR \cite{Tiusan2007,Faure-Vincent,Fanny}. Fig. 7b illustrates the phase shift image corresponding to the real space image of Fig. 7a showing the presence of the misfit dislocations inside the thick MgO barrier grown on (100) CFA . The density of dislocations is significantly larger than the one corresponding to the MgO barrier grown on (100) Fe (Fig. 7d). Consequently, as sketched in Fig. 7e and Fig. 7f, no coherent tunneling channel with conservation of $k_{\parallel}=0$ (denoted here by CH1) can be defined in the MgO barriers grown on CFA, whereas in the case of the thinner MgO barrier grown on Fe coherent tunneling channels are present. Therefore, the tunneling transport in the case of the thick MgO barrier will be dominated by incoherent hopping tunneling channels (denoted CH2) associated with dislocations and lattice deformations (Fig.7b and Fig. 3d) within the thick MgO barrier. These incoherent hopping transport channels are temperature activated and, as we demonstrated from extended GM analysis, they can explain the strong temperature dependence of the conductance and  TMR in our thick MgO barrier CFA based MTJs. The signature of the incoherent inelastic transport channels is also suggested by the parabolic dynamic conductance curves (Fig. 7g), similar to those measured in MTJs with polycrystalline or amorphous tunnel barriers. This also demonstrates that even in MTJs with crystalline electrodes, when the structural coherency is lost across the stack and coherent direct tunneling is quenched detrimental to incoherent hopping assisted transport, the fine signature of electrode density of states becomes smeared within the parabolic dominant shape of the dI/dV curves. The standard free-electron model applies perfectly (not shown here) and the analysis using the Brinkman model \cite{Brinkman} gives barrier parameters in agreement to those obtained from HRTEM analysis.

\begin{figure}
\begin{center}
\includegraphics[width=1\linewidth]{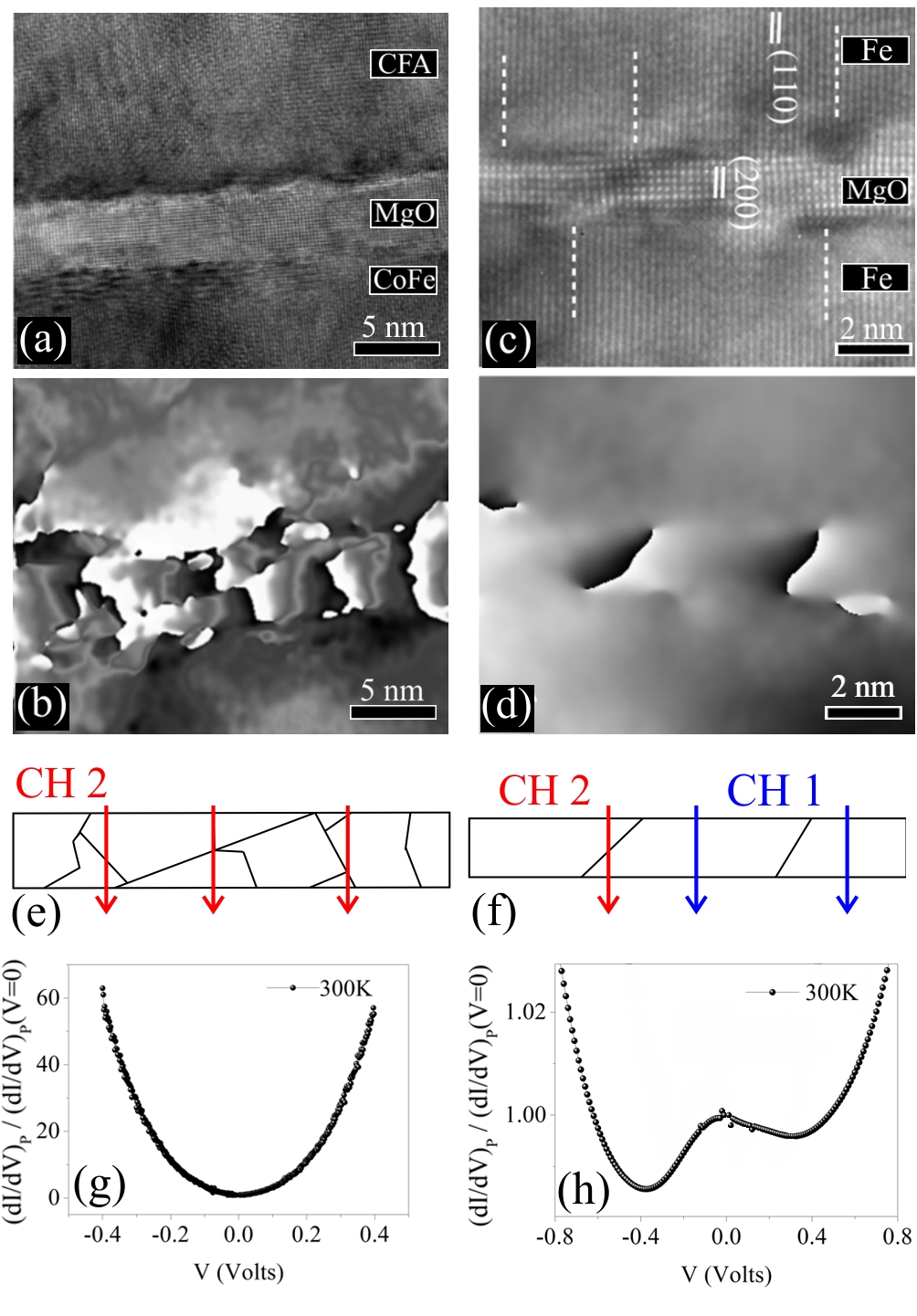}
\caption{\label{fig5} (a) HRTEM image of the CFA/MgO/CoFe trilayer and  (b) corresponding phase image; (c) HRTEM image of the Fe/MgO/Fe trilayer and (d) corresponding phase image; (e) and (f) schematic representation of the misfit dislocations inside the MgO barrier as resulted from (b) and (d), the coherent (CH1) and incoherent (CH2) conduction channels are also schematized; (g) and (h) bias voltage dependence of differential tunneling conductance dI/dV curves for the CFA/MgO/CoFe  and Fe/MgO/Fe MTJs, respectively.}
\end{center}
\end{figure}

The analysis performed on CFA/MgO/CoFe MTJs has been confronted to similar analysis performed on standard Fe/MgO/Fe MTJs known to provide efficient spin and symmetry filtering \cite{Tiusan2007,Faure-Vincent,Fanny}. In contrast to CFA/MgO/CoFe MTJs, one can clearly observe (Fig. 7d,f) that in case of the MgO grown on Fe, the lower corresponding period of misfit dislocation will allow a significant density of direct coherent tunneling channels (CH2). The tunneling spectroscopy experiments in these control systems demonstrate the coherent tunneling with k conservation (enabled by structural coherence across the stack) via the complex features in the dI/dV (V) curves demonstrating spin and symmetry filtering effects. As we previously argued \cite{Tiusan2007}, the presence of local minima event at RT in G$_{P}$(T) measured in Fe/MgO/Fe MTJs validates the symmetry filtering effects by the activation of additional $\Delta_5$ symmetry channel (Fig. 7h). \cite{Faure-Vincent, Tiusan2007}.

In order to compare with the case of CFA/MgO/CoFe MTJs where incoherent transport channels are dominant (Fig. 7e) and to give further proof of our hypothesis concerning the dominant direct coherent tunneling with respect to the spin-conserving hopping processes in Fe/MgO/Fe MTJ (Fig. 7f), we have performed the same transport in variable temperature analysis based on the extended Glazman-Matveev model. Figure 8a shows the dependence of the AP conductance and the corresponding fit within the GM model applied to $G_{AP}(T)$ curves measured in Fe/MgO/Fe MTJs. In contrast to CFA/MgO/CoFe MTJs, here the experimental data are correctly fited only implicating up to 3-LS chains. Figure 8b illustrates the relative contribution of each conduction channel to the total conductance in the AP state. Although 2 inelastic hopping conduction channels are thermally activated the 0-LS and 1-LS remains dominant even at room temperature, as expected from the HRTEM analysis and tunneling spectroscopy experiments. The fit of the TMR dependence on temperature (not shown here) gives an effective polarization of $P_0=0.84$, as expected for a (100)Fe electrode in conjunction with a (100)MgO barrier \cite{Parkin2004}, which benefits from $\Delta_1$ band symmetry filtering properties provided by the symmetry dependent attenuation rate of the barrier. Moreover, our analysis performed on Fe/MgO/Fe MTJs emphasizes the negative role of dislocations in these systems on the tunneling spin polarization amplitude. Even if larger TMR ratios are measured at RT in these systems, the value of the experimental TMR is lower that the theoretical predictions. The reduction of TMR has been previously related in literature to the negative effect of dislocations within the barrier on the symmetry filtering within the direct tunneling channels \cite{Tiusan2007}. However, our analysis developed here demonstrate an additional spin depolarizing mechanism in single crystal MTJs. This mechanism is related to spin-conserving hopping tunneling processes on localized levels associated to dislocations within the MgO barrier.

\begin{figure}
\begin{center}
\includegraphics[width=1\linewidth]{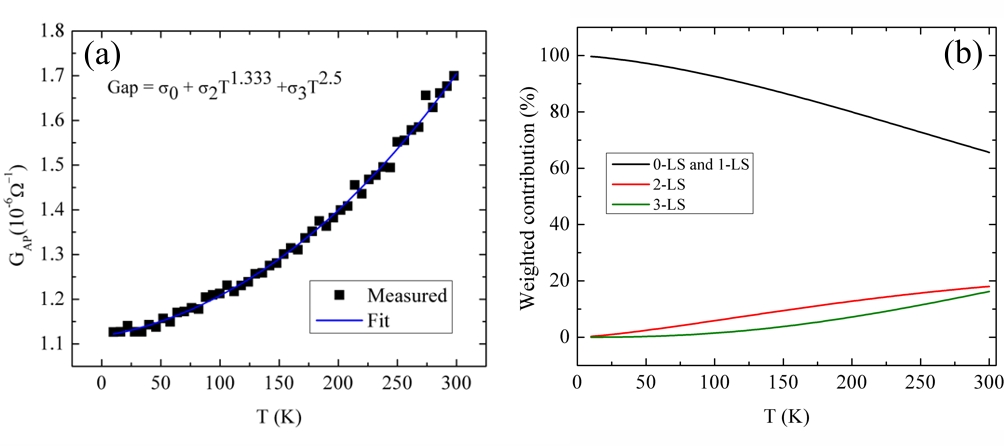}
\caption{\label{fig5} (a) Temperature dependence of the tunnel conductance in the AP state. (b) Relative contribution of the N-LS chains to the total conductance in the AP state.}
\end{center}
\end{figure}

In conclusion, our correlated structural-transport analysis in CFA/MgO and reference Fe/MgO MTJs demonstrates the negative role of structural extended defects in the MTJ stack on the tunneling polarization.
Related to the quenching of the crystal symmetry coherence across the MTJ stack, in these systems the coherence of the electronic Bloch functions in the single crystal CFA electrode is destroyed both along the propagation direction, by the oblique dislocation network within the thick barrier, and in the plane of the MTJ stack due to the large density of  misfit dislocations.  Moreover, this reduction of the lateral coherence of the wave function in single crystal MTJs is particularly detrimental to symmetry dependent filtering in single crystal MTJs, where the attenuation rate within the insulator depends on the in-plane modulation of the tunneling Bloch wave function \cite{Butler2001Red}. Our analysis provides a reasonable effective spin polarization of 0.57 at OK for our CFA/MgO-thick system which confirms the absence of $\Delta_1$ band symmetry filtering within the thick MgO barrier. Beyond of the direct effect on symmetry filtering with consequences on effective tunneling polarization amplitude, we demonstrate here that the dislocations may have a much larger negative impact on the tunneling magnetoresistance amplitude. In our Heusler based MTJs with thick MgO barrier, leading to large density of dislocations, the transport in variable temperature experiments fitted by theoretical modeling combined HRTEM phase analysis demonstrates the dominance of thermally activated inelastic hopping channels. This explains the relative low values of the TMR and its strong decrease in temperature. In contrast, in the case of reference Fe/MgO systems, where the coherent channels dominates, an effective spin polarization of 0.84 has been determined. This is an agreement with the large 180\% TMR measured in these systems at RT.  We can intuitively extrapolate our results and assume that even if longitudinal coherent channels (CH1) would exist (i.e. in CFA/MgO systems with thinner MgO barrier), a large density of dislocations and lattice deformations within the CFA films (Fig. 3c,d) and the barrier (Fig. 3d and 7b) would have a direct effect on the reduction of lateral coherence of the Bloch function within the electrodes. This could have negative effects on symmetry dependent tunneling in the single crystal MTJs.  Moreover, it would explain the relative few results reported in the literature concerning clear identification of symmetry dependent tunneling features \cite{Wang2010,Wang2010B}, even for the case of CFA/MgO MTJ systems showing significantly larger TMR ratios.

\begin{acknowledgments}

This work was partially supported by  POS CCE Project ID.574, code SMIS-CSNR 12467 and CNCSIS UEFISCSU Project No. PNII IDEI 4/2010 code ID-106.
C.T. acknowledges SPINCHAT Project No. ANR- 07-BLAN-341, B. Negulescu and G. Lengaigne for help in sample preparation and MTJ patterning.

\end{acknowledgments}

\bibliography{bib}

\providecommand{\noopsort}[1]{}\providecommand{\singleletter}[1]{#1}%
\begin{thebibliography}{25}%
\makeatletter
\providecommand \@ifxundefined [1]{%
 \@ifx{#1\undefined}
}%
\providecommand \@ifnum [1]{%
 \ifnum #1\expandafter \@firstoftwo
 \else \expandafter \@secondoftwo
 \fi
}%
\providecommand \@ifx [1]{%
 \ifx #1\expandafter \@firstoftwo
 \else \expandafter \@secondoftwo
 \fi
}%
\providecommand \natexlab [1]{#1}%
\providecommand \enquote  [1]{``#1''}%
\providecommand \bibnamefont  [1]{#1}%
\providecommand \bibfnamefont [1]{#1}%
\providecommand \citenamefont [1]{#1}%
\providecommand \href@noop [0]{\@secondoftwo}%
\providecommand \href [0]{\begingroup \@sanitize@url \@href}%
\providecommand \@href[1]{\@@startlink{#1}\@@href}%
\providecommand \@@href[1]{\endgroup#1\@@endlink}%
\providecommand \@sanitize@url [0]{\catcode `\\12\catcode `\$12\catcode
  `\&12\catcode `\#12\catcode `\^12\catcode `\_12\catcode `\%12\relax}%
\providecommand \@@startlink[1]{}%
\providecommand \@@endlink[0]{}%
\providecommand \url  [0]{\begingroup\@sanitize@url \@url }%
\providecommand \@url [1]{\endgroup\@href {#1}{\urlprefix }}%
\providecommand \urlprefix  [0]{URL }%
\providecommand \Eprint [0]{\href }%
\providecommand \doibase [0]{http://dx.doi.org/}%
\providecommand \selectlanguage [0]{\@gobble}%
\providecommand \bibinfo  [0]{\@secondoftwo}%
\providecommand \bibfield  [0]{\@secondoftwo}%
\providecommand \translation [1]{[#1]}%
\providecommand \BibitemOpen [0]{}%
\providecommand \bibitemStop [0]{}%
\providecommand \bibitemNoStop [0]{.\EOS\space}%
\providecommand \EOS [0]{\spacefactor3000\relax}%
\providecommand \BibitemShut  [1]{\csname bibitem#1\endcsname}%
\let\auto@bib@innerbib\@empty
\bibitem [{\citenamefont {Yuasa}\ \emph {et~al.}(2004)\citenamefont {Yuasa},
  \citenamefont {Nagahama}, \citenamefont {Fukushima}, \citenamefont {Suzuki},\
  and\ \citenamefont {Ando}}]{Yuasa-NatMat2004}%
  \BibitemOpen
  \bibfield  {author} {\bibinfo {author} {\bibfnamefont {S.}~\bibnamefont
  {Yuasa}}, \bibinfo {author} {\bibfnamefont {T.}~\bibnamefont {Nagahama}},
  \bibinfo {author} {\bibfnamefont {A.}~\bibnamefont {Fukushima}}, \bibinfo
  {author} {\bibfnamefont {Y.}~\bibnamefont {Suzuki}}, \ and\ \bibinfo {author}
  {\bibfnamefont {K.}~\bibnamefont {Ando}},\ }\href {\doibase
  doi:10.1038/nmat1257} {\bibfield  {journal} {\bibinfo  {journal} {Nature
  Materials}\ }\textbf {\bibinfo {volume} {3}},\ \bibinfo {pages} {868}
  (\bibinfo {year} {2004})}\BibitemShut {NoStop}%
\bibitem [{\citenamefont {Parkin}\ \emph {et~al.}(2004)\citenamefont {Parkin},
  \citenamefont {Kaiser}, \citenamefont {Panchula}, \citenamefont {Rice},
  \citenamefont {Hughes}, \citenamefont {Samant},\ and\ \citenamefont
  {Yang}}]{Parkin2004}%
  \BibitemOpen
  \bibfield  {author} {\bibinfo {author} {\bibfnamefont {S.~S.~P.}\
  \bibnamefont {Parkin}}, \bibinfo {author} {\bibfnamefont {C.}~\bibnamefont
  {Kaiser}}, \bibinfo {author} {\bibfnamefont {A.}~\bibnamefont {Panchula}},
  \bibinfo {author} {\bibfnamefont {P.~M.}\ \bibnamefont {Rice}}, \bibinfo
  {author} {\bibfnamefont {B.}~\bibnamefont {Hughes}}, \bibinfo {author}
  {\bibfnamefont {M.}~\bibnamefont {Samant}}, \ and\ \bibinfo {author}
  {\bibfnamefont {S.-H.}\ \bibnamefont {Yang}},\ }\href {\doibase
  10.1038/nmat1256} {\bibfield  {journal} {\bibinfo  {journal} {Nature
  Materials}\ }\textbf {\bibinfo {volume} {3}},\ \bibinfo {pages} {862}
  (\bibinfo {year} {2004})}\BibitemShut {NoStop}%
\bibitem [{\citenamefont {Tsunekawa}\ \emph {et~al.}(2005)\citenamefont
  {Tsunekawa}, \citenamefont {Djayaprawira}, \citenamefont {Nagai},
  \citenamefont {Maehara}, \citenamefont {Yamagata}, \citenamefont {Watanabe},
  \citenamefont {Yuasa}, \citenamefont {Suzuki},\ and\ \citenamefont
  {Ando}}]{Djayaprawira2005}%
  \BibitemOpen
  \bibfield  {author} {\bibinfo {author} {\bibfnamefont {K.}~\bibnamefont
  {Tsunekawa}}, \bibinfo {author} {\bibfnamefont {D.~D.}\ \bibnamefont
  {Djayaprawira}}, \bibinfo {author} {\bibfnamefont {M.}~\bibnamefont {Nagai}},
  \bibinfo {author} {\bibfnamefont {H.}~\bibnamefont {Maehara}}, \bibinfo
  {author} {\bibfnamefont {S.}~\bibnamefont {Yamagata}}, \bibinfo {author}
  {\bibfnamefont {N.}~\bibnamefont {Watanabe}}, \bibinfo {author}
  {\bibfnamefont {S.}~\bibnamefont {Yuasa}}, \bibinfo {author} {\bibfnamefont
  {Y.}~\bibnamefont {Suzuki}}, \ and\ \bibinfo {author} {\bibfnamefont
  {K.}~\bibnamefont {Ando}},\ }\href {\doibase DOI:10.1063/1.2012525}
  {\bibfield  {journal} {\bibinfo  {journal} {Appl. Phys. Lett.}\ }\textbf
  {\bibinfo {volume} {87}},\ \bibinfo {pages} {072503} (\bibinfo {year}
  {2005})}\BibitemShut {NoStop}%
\bibitem [{\citenamefont {Mathon}\ and\ \citenamefont
  {Umerski}(2001)}]{Mathon2001}%
  \BibitemOpen
  \bibfield  {author} {\bibinfo {author} {\bibfnamefont {J.}~\bibnamefont
  {Mathon}}\ and\ \bibinfo {author} {\bibfnamefont {A.}~\bibnamefont
  {Umerski}},\ }\href {\doibase 10.1103/PhysRevB.63.220403} {\bibfield
  {journal} {\bibinfo  {journal} {Phys. Rev. B}\ }\textbf {\bibinfo {volume}
  {63}},\ \bibinfo {pages} {220403} (\bibinfo {year} {2001})}\BibitemShut
  {NoStop}%
\bibitem [{\citenamefont {Butler}\ \emph
  {et~al.}(2001{\natexlab{a}})\citenamefont {Butler}, \citenamefont {Zhang},
  \citenamefont {Schulthess},\ and\ \citenamefont {MacLaren}}]{Butler2001}%
  \BibitemOpen
  \bibfield  {author} {\bibinfo {author} {\bibfnamefont {W.~H.}\ \bibnamefont
  {Butler}}, \bibinfo {author} {\bibfnamefont {X.-G.}\ \bibnamefont {Zhang}},
  \bibinfo {author} {\bibfnamefont {T.~C.}\ \bibnamefont {Schulthess}}, \ and\
  \bibinfo {author} {\bibfnamefont {J.~M.}\ \bibnamefont {MacLaren}},\ }\href
  {\doibase 10.1103/PhysRevB.63.054416} {\bibfield  {journal} {\bibinfo
  {journal} {Phys. Rev. B}\ }\textbf {\bibinfo {volume} {63}},\ \bibinfo
  {pages} {054416} (\bibinfo {year} {2001}{\natexlab{a}})}\BibitemShut
  {NoStop}%
\bibitem [{\citenamefont {Galanakis}\ \emph {et~al.}(2002)\citenamefont
  {Galanakis}, \citenamefont {Dederichs},\ and\ \citenamefont
  {Papanikolaou}}]{Galanakis2002}%
  \BibitemOpen
  \bibfield  {author} {\bibinfo {author} {\bibfnamefont {I.}~\bibnamefont
  {Galanakis}}, \bibinfo {author} {\bibfnamefont {P.~H.}\ \bibnamefont
  {Dederichs}}, \ and\ \bibinfo {author} {\bibfnamefont {N.}~\bibnamefont
  {Papanikolaou}},\ }\href {\doibase 10.1103/PhysRevB.66.174429} {\bibfield
  {journal} {\bibinfo  {journal} {Phys. Rev. B}\ }\textbf {\bibinfo {volume}
  {66}},\ \bibinfo {pages} {174429} (\bibinfo {year} {2002})}\BibitemShut
  {NoStop}%
\bibitem [{\citenamefont {Picozzi}\ \emph {et~al.}(2002)\citenamefont
  {Picozzi}, \citenamefont {Continenza},\ and\ \citenamefont
  {Freeman}}]{Picozzi2002}%
  \BibitemOpen
  \bibfield  {author} {\bibinfo {author} {\bibfnamefont {S.}~\bibnamefont
  {Picozzi}}, \bibinfo {author} {\bibfnamefont {A.}~\bibnamefont {Continenza}},
  \ and\ \bibinfo {author} {\bibfnamefont {A.~J.}\ \bibnamefont {Freeman}},\
  }\href {\doibase 10.1103/PhysRevB.66.094421} {\bibfield  {journal} {\bibinfo
  {journal} {Phys. Rev. B}\ }\textbf {\bibinfo {volume} {66}},\ \bibinfo
  {pages} {094421} (\bibinfo {year} {2002})}\BibitemShut {NoStop}%
\bibitem [{\citenamefont {Wang}\ \emph {et~al.}(2009)\citenamefont {Wang},
  \citenamefont {Sukegawa}, \citenamefont {Shan}, \citenamefont {Mitani},\ and\
  \citenamefont {Inomata}}]{Wang2009}%
  \BibitemOpen
  \bibfield  {author} {\bibinfo {author} {\bibfnamefont {W.}~\bibnamefont
  {Wang}}, \bibinfo {author} {\bibfnamefont {H.}~\bibnamefont {Sukegawa}},
  \bibinfo {author} {\bibfnamefont {R.}~\bibnamefont {Shan}}, \bibinfo {author}
  {\bibfnamefont {S.}~\bibnamefont {Mitani}}, \ and\ \bibinfo {author}
  {\bibfnamefont {K.}~\bibnamefont {Inomata}},\ }\href {\doibase
  DOI:10.1063/1.3258069} {\bibfield  {journal} {\bibinfo  {journal} {Appl.
  Phys. Lett.}\ }\textbf {\bibinfo {volume} {95}},\ \bibinfo {pages} {182502}
  (\bibinfo {year} {2009})}\BibitemShut {NoStop}%
\bibitem [{\citenamefont {Wang}\ \emph
  {et~al.}(2010{\natexlab{a}})\citenamefont {Wang}, \citenamefont {Liu},
  \citenamefont {Kodzuka}, \citenamefont {Sukegawa}, \citenamefont {Wojcik},
  \citenamefont {Jedryka}, \citenamefont {Wu}, \citenamefont {Inomata},
  \citenamefont {Mitani},\ and\ \citenamefont {Hono}}]{Wang2010}%
  \BibitemOpen
  \bibfield  {author} {\bibinfo {author} {\bibfnamefont {W.}~\bibnamefont
  {Wang}}, \bibinfo {author} {\bibfnamefont {E.}~\bibnamefont {Liu}}, \bibinfo
  {author} {\bibfnamefont {M.}~\bibnamefont {Kodzuka}}, \bibinfo {author}
  {\bibfnamefont {H.}~\bibnamefont {Sukegawa}}, \bibinfo {author}
  {\bibfnamefont {M.}~\bibnamefont {Wojcik}}, \bibinfo {author} {\bibfnamefont
  {E.}~\bibnamefont {Jedryka}}, \bibinfo {author} {\bibfnamefont {G.~H.}\
  \bibnamefont {Wu}}, \bibinfo {author} {\bibfnamefont {K.}~\bibnamefont
  {Inomata}}, \bibinfo {author} {\bibfnamefont {S.}~\bibnamefont {Mitani}}, \
  and\ \bibinfo {author} {\bibfnamefont {K.}~\bibnamefont {Hono}},\ }\href
  {\doibase 10.1103/PhysRevB.81.140402} {\bibfield  {journal} {\bibinfo
  {journal} {Phys. Rev. B}\ }\textbf {\bibinfo {volume} {81}},\ \bibinfo
  {pages} {140402} (\bibinfo {year} {2010}{\natexlab{a}})}\BibitemShut
  {NoStop}%
\bibitem [{\citenamefont {Wang}\ \emph
  {et~al.}(2010{\natexlab{b}})\citenamefont {Wang}, \citenamefont {Sukegawa},\
  and\ \citenamefont {Inomata}}]{Wang2010B}%
  \BibitemOpen
  \bibfield  {author} {\bibinfo {author} {\bibfnamefont {W.}~\bibnamefont
  {Wang}}, \bibinfo {author} {\bibfnamefont {H.}~\bibnamefont {Sukegawa}}, \
  and\ \bibinfo {author} {\bibfnamefont {K.}~\bibnamefont {Inomata}},\ }\href
  {\doibase 10.1103/PhysRevB.82.092402} {\bibfield  {journal} {\bibinfo
  {journal} {Phys. Rev. B}\ }\textbf {\bibinfo {volume} {82}},\ \bibinfo
  {pages} {092402} (\bibinfo {year} {2010}{\natexlab{b}})}\BibitemShut
  {NoStop}%
\bibitem [{\citenamefont {Wen}\ \emph {et~al.}(2011)\citenamefont {Wen},
  \citenamefont {Sukegawa}, \citenamefont {Mitani},\ and\ \citenamefont
  {Inomata}}]{texturedCFA}%
  \BibitemOpen
  \bibfield  {author} {\bibinfo {author} {\bibfnamefont {Z.}~\bibnamefont
  {Wen}}, \bibinfo {author} {\bibfnamefont {H.}~\bibnamefont {Sukegawa}},
  \bibinfo {author} {\bibfnamefont {S.}~\bibnamefont {Mitani}}, \ and\ \bibinfo
  {author} {\bibfnamefont {K.}~\bibnamefont {Inomata}},\ }\href {\doibase
  DOI:10.1063/1.3587640} {\bibfield  {journal} {\bibinfo  {journal} {Appl.
  Phys. Lett.}\ }\textbf {\bibinfo {volume} {98}},\ \bibinfo {pages} {192505}
  (\bibinfo {year} {2011})}\BibitemShut {NoStop}%
\bibitem [{\citenamefont {Mizukami}\ \emph {et~al.}(2009)\citenamefont
  {Mizukami}, \citenamefont {Watanabe}, \citenamefont {Oogane}, \citenamefont
  {Ando}, \citenamefont {Miura}, \citenamefont {Shirai},\ and\ \citenamefont
  {Miyazaki}}]{Mizukami2009}%
  \BibitemOpen
  \bibfield  {author} {\bibinfo {author} {\bibfnamefont {S.}~\bibnamefont
  {Mizukami}}, \bibinfo {author} {\bibfnamefont {D.}~\bibnamefont {Watanabe}},
  \bibinfo {author} {\bibfnamefont {M.}~\bibnamefont {Oogane}}, \bibinfo
  {author} {\bibfnamefont {Y.}~\bibnamefont {Ando}}, \bibinfo {author}
  {\bibfnamefont {Y.}~\bibnamefont {Miura}}, \bibinfo {author} {\bibfnamefont
  {M.}~\bibnamefont {Shirai}}, \ and\ \bibinfo {author} {\bibfnamefont
  {T.}~\bibnamefont {Miyazaki}},\ }\href {\doibase DOI:10.1063/1.3067607}
  {\bibfield  {journal} {\bibinfo  {journal} {J. Appl. Phys.}\ }\textbf
  {\bibinfo {volume} {105}},\ \bibinfo {pages} {07D306} (\bibinfo {year}
  {2009})}\BibitemShut {NoStop}%
\bibitem [{\citenamefont {Ortiz}\ \emph {et~al.}(2011)\citenamefont {Ortiz},
  \citenamefont {Gabor}, \citenamefont {Jr}, \citenamefont {Boust},
  \citenamefont {Issac}, \citenamefont {Tiusan}, \citenamefont {Hehn},\ and\
  \citenamefont {Bobo}}]{GaborJAP}%
  \BibitemOpen
  \bibfield  {author} {\bibinfo {author} {\bibfnamefont {G.}~\bibnamefont
  {Ortiz}}, \bibinfo {author} {\bibfnamefont {M.~S.}\ \bibnamefont {Gabor}},
  \bibinfo {author} {\bibfnamefont {T.~P.}\ \bibnamefont {Jr}}, \bibinfo
  {author} {\bibfnamefont {F.}~\bibnamefont {Boust}}, \bibinfo {author}
  {\bibfnamefont {F.}~\bibnamefont {Issac}}, \bibinfo {author} {\bibfnamefont
  {C.}~\bibnamefont {Tiusan}}, \bibinfo {author} {\bibfnamefont
  {M.}~\bibnamefont {Hehn}}, \ and\ \bibinfo {author} {\bibfnamefont {J.~F.}\
  \bibnamefont {Bobo}},\ }\href {\doibase DOI:10.1063/1.3549581} {\bibfield
  {journal} {\bibinfo  {journal} {J. Appl. Phys.}\ }\textbf {\bibinfo {volume}
  {109}},\ \bibinfo {pages} {07D324} (\bibinfo {year} {2011})}\BibitemShut
  {NoStop}%
\bibitem [{\citenamefont {Hytch}\ \emph {et~al.}(1998)\citenamefont {Hytch},
  \citenamefont {Snoeck},\ and\ \citenamefont {Kilaas}}]{Snoeck}%
  \BibitemOpen
  \bibfield  {author} {\bibinfo {author} {\bibfnamefont {M.}~\bibnamefont
  {Hytch}}, \bibinfo {author} {\bibfnamefont {E.}~\bibnamefont {Snoeck}}, \
  and\ \bibinfo {author} {\bibfnamefont {R.}~\bibnamefont {Kilaas}},\ }\href
  {\doibase 10.1016/S0304-3991(98)00035-7} {\bibfield  {journal} {\bibinfo
  {journal} {Ultramicroscopy}\ }\textbf {\bibinfo {volume} {74}},\ \bibinfo
  {pages} {131 } (\bibinfo {year} {1998})}\BibitemShut {NoStop}%
\bibitem [{\citenamefont {Lu}\ \emph {et~al.}(2009)\citenamefont {Lu},
  \citenamefont {Tran}, \citenamefont {Jaffr\`es}, \citenamefont {Seneor},
  \citenamefont {Deranlot}, \citenamefont {Petroff}, \citenamefont {George},
  \citenamefont {L\'epine}, \citenamefont {Ababou},\ and\ \citenamefont
  {J\'ez\'equel}}]{Lu}%
  \BibitemOpen
  \bibfield  {author} {\bibinfo {author} {\bibfnamefont {Y.}~\bibnamefont
  {Lu}}, \bibinfo {author} {\bibfnamefont {M.}~\bibnamefont {Tran}}, \bibinfo
  {author} {\bibfnamefont {H.}~\bibnamefont {Jaffr\`es}}, \bibinfo {author}
  {\bibfnamefont {P.}~\bibnamefont {Seneor}}, \bibinfo {author} {\bibfnamefont
  {C.}~\bibnamefont {Deranlot}}, \bibinfo {author} {\bibfnamefont
  {F.}~\bibnamefont {Petroff}}, \bibinfo {author} {\bibfnamefont {J.-M.}\
  \bibnamefont {George}}, \bibinfo {author} {\bibfnamefont {B.}~\bibnamefont
  {L\'epine}}, \bibinfo {author} {\bibfnamefont {S.}~\bibnamefont {Ababou}}, \
  and\ \bibinfo {author} {\bibfnamefont {G.}~\bibnamefont {J\'ez\'equel}},\
  }\href {\doibase 10.1103/PhysRevLett.102.176801} {\bibfield  {journal}
  {\bibinfo  {journal} {Phys. Rev. Lett.}\ }\textbf {\bibinfo {volume} {102}},\
  \bibinfo {pages} {176801} (\bibinfo {year} {2009})}\BibitemShut {NoStop}%
\bibitem [{\citenamefont {Gabor}\ \emph {et~al.}(2011)\citenamefont {Gabor},
  \citenamefont {Petrisor~Jr.}, \citenamefont {Tiusan}, \citenamefont {Hehn},\
  and\ \citenamefont {Petrisor}}]{GaborPRB}%
  \BibitemOpen
  \bibfield  {author} {\bibinfo {author} {\bibfnamefont {M.~S.}\ \bibnamefont
  {Gabor}}, \bibinfo {author} {\bibfnamefont {T.}~\bibnamefont {Petrisor~Jr.}},
  \bibinfo {author} {\bibfnamefont {C.}~\bibnamefont {Tiusan}}, \bibinfo
  {author} {\bibfnamefont {M.}~\bibnamefont {Hehn}}, \ and\ \bibinfo {author}
  {\bibfnamefont {T.}~\bibnamefont {Petrisor}},\ }\href {\doibase
  10.1103/PhysRevB.84.134413} {\bibfield  {journal} {\bibinfo  {journal} {Phys.
  Rev. B}\ }\textbf {\bibinfo {volume} {84}},\ \bibinfo {pages} {134413}
  (\bibinfo {year} {2011})}\BibitemShut {NoStop}%
\bibitem [{\citenamefont {Faure-Vincent}\ \emph {et~al.}(2003)\citenamefont
  {Faure-Vincent}, \citenamefont {Tiusan}, \citenamefont {Jouguelet},
  \citenamefont {Canet}, \citenamefont {Sajieddine}, \citenamefont {Bellouard},
  \citenamefont {Popova}, \citenamefont {Hehn}, \citenamefont {Montaigne},\
  and\ \citenamefont {Schuhl}}]{Faure-Vincent}%
  \BibitemOpen
  \bibfield  {author} {\bibinfo {author} {\bibfnamefont {J.}~\bibnamefont
  {Faure-Vincent}}, \bibinfo {author} {\bibfnamefont {C.}~\bibnamefont
  {Tiusan}}, \bibinfo {author} {\bibfnamefont {E.}~\bibnamefont {Jouguelet}},
  \bibinfo {author} {\bibfnamefont {F.}~\bibnamefont {Canet}}, \bibinfo
  {author} {\bibfnamefont {M.}~\bibnamefont {Sajieddine}}, \bibinfo {author}
  {\bibfnamefont {C.}~\bibnamefont {Bellouard}}, \bibinfo {author}
  {\bibfnamefont {E.}~\bibnamefont {Popova}}, \bibinfo {author} {\bibfnamefont
  {M.}~\bibnamefont {Hehn}}, \bibinfo {author} {\bibfnamefont {F.}~\bibnamefont
  {Montaigne}}, \ and\ \bibinfo {author} {\bibfnamefont {A.}~\bibnamefont
  {Schuhl}},\ }\href {\doibase DOI:10.1063/1.1586785} {\bibfield  {journal}
  {\bibinfo  {journal} {Appl. Phys. Lett.}\ }\textbf {\bibinfo {volume} {82}},\
  \bibinfo {pages} {4507} (\bibinfo {year} {2003})}\BibitemShut {NoStop}%
\bibitem [{\citenamefont {Shang}\ \emph {et~al.}(1998)\citenamefont {Shang},
  \citenamefont {Nowak}, \citenamefont {Jansen},\ and\ \citenamefont
  {Moodera}}]{Shang98}%
  \BibitemOpen
  \bibfield  {author} {\bibinfo {author} {\bibfnamefont {C.~H.}\ \bibnamefont
  {Shang}}, \bibinfo {author} {\bibfnamefont {J.}~\bibnamefont {Nowak}},
  \bibinfo {author} {\bibfnamefont {R.}~\bibnamefont {Jansen}}, \ and\ \bibinfo
  {author} {\bibfnamefont {J.~S.}\ \bibnamefont {Moodera}},\ }\href {\doibase
  10.1103/PhysRevB.58.R2917} {\bibfield  {journal} {\bibinfo  {journal} {Phys.
  Rev. B}\ }\textbf {\bibinfo {volume} {58}},\ \bibinfo {pages} {R2917}
  (\bibinfo {year} {1998})}\BibitemShut {NoStop}%
\bibitem [{\citenamefont {Miura}\ \emph {et~al.}(2004)\citenamefont {Miura},
  \citenamefont {Nagao},\ and\ \citenamefont {Shirai}}]{Miura04}%
  \BibitemOpen
  \bibfield  {author} {\bibinfo {author} {\bibfnamefont {Y.}~\bibnamefont
  {Miura}}, \bibinfo {author} {\bibfnamefont {K.}~\bibnamefont {Nagao}}, \ and\
  \bibinfo {author} {\bibfnamefont {M.}~\bibnamefont {Shirai}},\ }\href
  {\doibase 10.1103/PhysRevB.69.144413} {\bibfield  {journal} {\bibinfo
  {journal} {Phys. Rev. B}\ }\textbf {\bibinfo {volume} {69}},\ \bibinfo
  {pages} {144413} (\bibinfo {year} {2004})}\BibitemShut {NoStop}%
\bibitem [{\citenamefont {Inomata}\ \emph {et~al.}(2008)\citenamefont
  {Inomata}, \citenamefont {Ikeda}, \citenamefont {Tezuka}, \citenamefont
  {Goto}, \citenamefont {Sugimoto}, \citenamefont {Wojcik},\ and\ \citenamefont
  {Jedryka}}]{Inomata08}%
  \BibitemOpen
  \bibfield  {author} {\bibinfo {author} {\bibfnamefont {K.}~\bibnamefont
  {Inomata}}, \bibinfo {author} {\bibfnamefont {N.}~\bibnamefont {Ikeda}},
  \bibinfo {author} {\bibfnamefont {N.}~\bibnamefont {Tezuka}}, \bibinfo
  {author} {\bibfnamefont {R.}~\bibnamefont {Goto}}, \bibinfo {author}
  {\bibfnamefont {S.}~\bibnamefont {Sugimoto}}, \bibinfo {author}
  {\bibfnamefont {M.}~\bibnamefont {Wojcik}}, \ and\ \bibinfo {author}
  {\bibfnamefont {E.}~\bibnamefont {Jedryka}},\ }\href
  {http://stacks.iop.org/1468-6996/9/i=1/a=014101} {\bibfield  {journal}
  {\bibinfo  {journal} {Science and Technology of Advanced Materials}\ }\textbf
  {\bibinfo {volume} {9}},\ \bibinfo {pages} {014101} (\bibinfo {year}
  {2008})}\BibitemShut {NoStop}%
\bibitem [{\citenamefont {Schebaum}\ \emph {et~al.}(2010)\citenamefont
  {Schebaum}, \citenamefont {Ebke}, \citenamefont {Niemeyer}, \citenamefont
  {Reiss}, \citenamefont {Moodera},\ and\ \citenamefont {Thomas}}]{schebaum}%
  \BibitemOpen
  \bibfield  {author} {\bibinfo {author} {\bibfnamefont {O.}~\bibnamefont
  {Schebaum}}, \bibinfo {author} {\bibfnamefont {D.}~\bibnamefont {Ebke}},
  \bibinfo {author} {\bibfnamefont {A.}~\bibnamefont {Niemeyer}}, \bibinfo
  {author} {\bibfnamefont {G.}~\bibnamefont {Reiss}}, \bibinfo {author}
  {\bibfnamefont {J.~S.}\ \bibnamefont {Moodera}}, \ and\ \bibinfo {author}
  {\bibfnamefont {A.}~\bibnamefont {Thomas}},\ }\href {\doibase
  10.1063/1.3358245} {\bibfield  {journal} {\bibinfo  {journal} {Journal of
  Applied Physics}\ }\textbf {\bibinfo {volume} {107}},\ \bibinfo {eid}
  {09C717} (\bibinfo {year} {2010})}\BibitemShut {NoStop}%
\bibitem [{\citenamefont {Tiusan}\ \emph {et~al.}(2007)\citenamefont {Tiusan},
  \citenamefont {Greullet}, \citenamefont {Hehn}, \citenamefont {Montaigne},
  \citenamefont {Andrieu},\ and\ \citenamefont {Schuhl}}]{Tiusan2007}%
  \BibitemOpen
  \bibfield  {author} {\bibinfo {author} {\bibfnamefont {C.}~\bibnamefont
  {Tiusan}}, \bibinfo {author} {\bibfnamefont {F.}~\bibnamefont {Greullet}},
  \bibinfo {author} {\bibfnamefont {M.}~\bibnamefont {Hehn}}, \bibinfo {author}
  {\bibfnamefont {F.}~\bibnamefont {Montaigne}}, \bibinfo {author}
  {\bibfnamefont {S.}~\bibnamefont {Andrieu}}, \ and\ \bibinfo {author}
  {\bibfnamefont {A.}~\bibnamefont {Schuhl}},\ }\href@noop {} {\bibfield
  {journal} {\bibinfo  {journal} {J. of Phys.: Cond. Matt.}\ }\textbf {\bibinfo
  {volume} {19}},\ \bibinfo {pages} {165201} (\bibinfo {year}
  {2007})}\BibitemShut {NoStop}%
\bibitem [{\citenamefont {Greullet}\ \emph {et~al.}(2007)\citenamefont
  {Greullet}, \citenamefont {Tiusan}, \citenamefont {Montaigne}, \citenamefont
  {Hehn}, \citenamefont {Halley}, \citenamefont {Bengone}, \citenamefont
  {Bowen},\ and\ \citenamefont {Weber}}]{Fanny}%
  \BibitemOpen
  \bibfield  {author} {\bibinfo {author} {\bibfnamefont {F.}~\bibnamefont
  {Greullet}}, \bibinfo {author} {\bibfnamefont {C.}~\bibnamefont {Tiusan}},
  \bibinfo {author} {\bibfnamefont {F.}~\bibnamefont {Montaigne}}, \bibinfo
  {author} {\bibfnamefont {M.}~\bibnamefont {Hehn}}, \bibinfo {author}
  {\bibfnamefont {D.}~\bibnamefont {Halley}}, \bibinfo {author} {\bibfnamefont
  {O.}~\bibnamefont {Bengone}}, \bibinfo {author} {\bibfnamefont
  {M.}~\bibnamefont {Bowen}}, \ and\ \bibinfo {author} {\bibfnamefont
  {W.}~\bibnamefont {Weber}},\ }\href {\doibase 10.1103/PhysRevLett.99.187202}
  {\bibfield  {journal} {\bibinfo  {journal} {Phys. Rev. Lett.}\ }\textbf
  {\bibinfo {volume} {99}},\ \bibinfo {pages} {187202} (\bibinfo {year}
  {2007})}\BibitemShut {NoStop}%
\bibitem [{\citenamefont {Brinkman}\ \emph {et~al.}(1970)\citenamefont
  {Brinkman}, \citenamefont {Dynes},\ and\ \citenamefont {Rowell}}]{Brinkman}%
  \BibitemOpen
  \bibfield  {author} {\bibinfo {author} {\bibfnamefont {W.~F.}\ \bibnamefont
  {Brinkman}}, \bibinfo {author} {\bibfnamefont {R.~C.}\ \bibnamefont {Dynes}},
  \ and\ \bibinfo {author} {\bibfnamefont {J.~M.}\ \bibnamefont {Rowell}},\
  }\href {\doibase DOI:10.1063/1.1659141} {\bibfield  {journal} {\bibinfo
  {journal} {J. Appl. Phys.}\ }\textbf {\bibinfo {volume} {41}},\ \bibinfo
  {pages} {1915} (\bibinfo {year} {1970})}\BibitemShut {NoStop}%
\bibitem [{\citenamefont {Butler}\ \emph
  {et~al.}(2001{\natexlab{b}})\citenamefont {Butler}, \citenamefont {Zhang},
  \citenamefont {Schulthess},\ and\ \citenamefont {MacLaren}}]{Butler2001Red}%
  \BibitemOpen
  \bibfield  {author} {\bibinfo {author} {\bibfnamefont {W.~H.}\ \bibnamefont
  {Butler}}, \bibinfo {author} {\bibfnamefont {X.-G.}\ \bibnamefont {Zhang}},
  \bibinfo {author} {\bibfnamefont {T.~C.}\ \bibnamefont {Schulthess}}, \ and\
  \bibinfo {author} {\bibfnamefont {J.~M.}\ \bibnamefont {MacLaren}},\ }\href
  {\doibase 10.1103/PhysRevB.63.092402} {\bibfield  {journal} {\bibinfo
  {journal} {Phys. Rev. B}\ }\textbf {\bibinfo {volume} {63}},\ \bibinfo
  {pages} {092402} (\bibinfo {year} {2001}{\natexlab{b}})}\BibitemShut
  {NoStop}%
\end{thebibliography}%

\end{document}